\documentstyle[preprint,aps]{revtex}
\input epsf
\tighten
\draft
\begin{document}
\title{Numerical Results for the Hubbard Model: Implications for 
the High ${\mathbf T_c}$ Pairing Mechanism}

\author{Douglas J.~Scalapino\thanks{djs@vulcan.physics.ucsb.edu}}

\address{\sl Department of Physics,
         University of California,
         Santa Barbara, CA 93106-9530 USA}

\author{S.R.~White\thanks{srwhite@uci.edu}}

\address{\sl Department of Physics,
         University of California,
         Irvine, CA 92697 USA}

\date{\today}
\maketitle

\vskip .26 in
\centerline{\bf Dedicated to Martin C.~Gutzwiller on the occasion of his
75$^{\rm th}$ birthday.}

\begin{abstract}

Numerical studies of the Hubbard model and its strong-coupling form, the
$t$-$J$ model, show evidence for antiferromagnetic, $d_{x^2-y^2}$-pairing
and stripe correlations which remind one of phenomena seen in the layered
cuprate materials.  Here, we ask what these numerical results imply 
about various scenarios for the pairing mechanism.

\end{abstract}

\vskip 2.75 in

\noindent

\setlength{\baselineskip}{.2in}
\renewcommand{\thefootnote}{\fnsymbol{footnote}}


In his 1963 article on the {\sl Effect of Correlation on the Ferromagnetism
of Transition Metals} \cite{Gut63}, Gutzwiller suggested that it was 
``instructive'' to
consider a model Hamiltonian of the form
\begin{equation}
H= \sum_{ks}\, \epsilon_k n_{ks} + U \sum_i\, n_{i\uparrow}n_{i\downarrow}\ .
\label{one}
\end{equation}
He noted that the first term arose from solving Schr\"odinger's equation
for one electron in the effective periodic potential of the lattice while
the second term described the Coulomb repulsion between two electrons of
opposite spin which happen to be in the same orbit at site $i$. In this
same year, in a paper entitled {\sl Electron Correlations in Narrow
Bands} \cite{Hub63},
Hubbard also studied this model.  He wrote $H$ in the form
\begin{equation}
H= -\sum_{ijs}\, t_{ij}\, \left(c^+_{i,s} c_{j,s} + c^+_{j,s}c_{i,s}\right)
+ U\sum_i\, n_{i\uparrow}n_{i\downarrow}
\label{two}
\end{equation}
which is now called the Hubbard model. Here $c^+_{i,s}$ created an electron
of spin $s$ in an orbital at site $i$ and $n_{is}=c^+_{i,s}c_{i,s}$ was the
orbital occupation number for electrons of spin $s$ on site $i$. For a 2D
square lattice with $t_{ij}=t$ for near-neighbor sites only, $\epsilon_k$ of
Eq.~(\ref{one}) is $-2t(\cos k_x + \cos\, k_y)$. In this case, two parameters,
$U/t$ and the average site occupation $n$ determine the state of the
system. In his 1963 paper, Gutzwiller also proposed a new trial wave
function for large $U/t$. This wave function was obtained by applying a
projection operator
\begin{equation}
P_G(\eta)= \mathop{\Pi}_i\, \left(1-(1-\eta)
\, n_{i\uparrow}n_{i\downarrow}\right)
\label{three}
\end{equation}
to the usual Slater determinant of Bloch functions.  In particular, for
$\eta=0$, $P_G(0)$ projects out all configurations with doubly occupied sites.
The $t$-$J$ model is the large $U/t$ limit of the Hubbard model obtained by
expanding to order $t^2/U$ and dropping three-site terms which are
proportional to the doping. Using the Gutzwiller projection operator, the
$t$-$J$ model has the form \cite{Hir85,GJR87,ZR88}
\begin{equation}
H=P_G (0)\left[-\sum_{\langle ij\rangle,s} t\, \left(c^+_{is}c_{js} +
c^+_{js}c_{is}\right) + J\sum_{\langle ij\rangle} \left(\vec S_i\cdot \vec
S_j - \frac{1}{4}\ n_in_j\right)\right] P_G (0)
\label{four}
\end{equation}
with the exchange interaction $J=4t^2/U$. Here the state of the system is
determined by $J/t$ and the site occupation $n$.

Both Gutzwiller and Hubbard were interested in the strongly-correlated case
in which $U/t$ was large and in understanding what conditions favored
ferromagnetism.  Now, almost forty years later, we are still seeking to
understand the behavior of the deceptively simple looking Hubbard and
$t$-$J$ models. In particular, we would like to know if these models
contain the essential physics of the high $T_c$ cuprate superconductors,
and if so what they tell us about the basic pairing mechanism. 
The premise of this brief report is that numerical calculations have shown
that the Hubbard and $t$-$J$ models exhibit a number of 
the phenomena which have been observed in various cuprate materials.
Nevertheless, as we will discuss, the question regarding the nature of the
underlying pairing mechanism remains open with a diverse range of proposals
set forth. 

Turning first to our premise, we know from Monte Carlo
calculations that the groundstate of the half-filled 2D Hubbard model has
long-range antiferromagnetic order \cite{Hir85a}
 as does the 2D Heisenberg model \cite{RY88}, which
is just the undoped $t$-$J$ model. It also has been shown from finite
temperature Monte Carlo calculations \cite{Bir99,Man91} that the low
energy spin-fluctuations of the insulating cuprates are well described by
the Heisenberg model.  In particular, the instantaneous spin-spin
correlation length calculated for the 2D $S=1/2$ Heisenberg model with a
near-neighbor exchange is in excellent agreement with neutron scattering
measurements of $La_2CuO_4$ above its Neel Temperature \cite{Bir99}.
Thus, the 2D Hubbard and $t$-$J$ models
provide an excellent description of the low energy magnetic behavior of the
undoped layered cuprates.  Likewise, DMRG calculations on half-filled
$n$-leg Hubbard and $t$-$J$ ladders have shown that the even-leg ladders are 
spin-gapped while the odd-leg ladders are not \cite{WNS94}. 
This is in agreement with 
experimental
data on $SrCu_2O_3$ and $Sr_2Cu_3O_5$ which contain weakly-coupled arrays of
2- and 3-leg ladders respectively \cite{AHTIK94}.

When a 2-leg Hubbard or $t$-$J$ ladder is doped away from half-filling, 
the added holes form
$d_{x^2-y^2}$-like pairs with power law pair-field and CDW correlations
and there are physically reasonable parameter regimes in which the
pair-field correlations are dominant \cite{DRS92,SRZ94,NWS94,DR96}.  
Under pressure, the 2-leg ladder
material $(SrCa)_{14}Cu_{24}O_{41}$ has a superconducting transition
\cite{Ueh96} with 
$T_c\simeq 12$K,
however transport evidence suggests that the pressure has made this
system into an anisotropic 2D material rather than a weakly Josephson
coupled array of 2-leg ladders. On doped 3-leg ladders, calculations
indicate that the holes 
first go into the odd band and for doping $x \lesssim.05$
 the system exhibits properties similar to
a 1-leg, $t$-$J$ ladder, which is known to have charge-spin separation
\cite{Ric97,WS98}. At higher doping in the 3-leg ladder,
pairs and domain walls form. When two holes are added to 4- and 6-leg
ladders, they form $d_{x^2-y^2}$-like pairs and at finite doping, domain
walls form. On doped 8-leg, $t$-$J$ ladders,
DMRG calculations \cite{WS98a} find domain walls with a linear hole 
density of 0.5 per
$Cu$ separating $\pi$-phase shifted antiferromagnetic regions, similar to
the stripes observed by neutron scattering \cite{Tra97} from
$La_{1.6-x}Nd_{0.4}Sr_xCuO_4$. When a small next-near-neighbor hopping
$t^\prime$ is added, $d_{x^2-y^2}$-pairing correlations are favored over
the striped domain wall configurations in the DMRG calculations
\cite{WS99}.

Nevertheless, considerable difference of opinion remains about
the nature of the correlations in the 2D system. While early
RPA calculations for the doped 2D Hubbard model found that the exchange of
antiferromagnetic spin fluctuations could lead to
$d_{x^2-y^2}$-superconductivity \cite{SLH86}, competing channels which arise 
from
nesting and van Hove singularities lead to a breakdown of conventional
perturbation theory.   A number of authors \cite{Met00} have carried out
renormalization group analyses which treat the various particle-hole and
particle-particle channels on an equal footing.  For the weak-coupling
doped Hubbard model with a near-neighbor hopping $t$, a 
$d_{x^2-y^2}$-superconducting state is inferred from the divergence of the 
one-loop
$d_{x^2-y^2}$-pairing susceptibility. More generally, the one-loop
calculation can lead to divergencies of vertices associated with various
channels. In particular, with a next-near-neighbor hopping $t^\prime$,
different instabilities arising from the position of Van Hove points on the
fermi surface can compete.  For example, from an analysis of the
renormalization group flow, Furukawa {\it et.~al.} \cite{FRS98}
 have recently suggested
that an Umklap-gapped, spin-liquid state might exist for a 2D Hubbard model
with appropriate values
of $t^\prime$ and doping.

The original finite temperature Monte Carlo calculations were unable to
reach low temperatures for the doped Hubbard model because of the fermion
determinantal sign problem.  They have, however, been used to study the
momentum and Matsubara frequency dependence of the effective pairing
interaction at temperatures of order one-half to one-third of the exchange
energy scale $4t^2/U$. The leading eigenvalue in the singlet channel
associated with this interaction was shown to have $d_{x^2-y^2}$ symmetry
\cite{BSW93}.
However, zero temperature constrained path
Monte Carlo calculations \cite{ZCG97} find only
short-range pairing correlations in the groundstate of the 2D Hubbard
model. 

For the 2D $t$-$J$ model, questions of phase separation, stripe formation,
and pairing arise.  Our study of long $n$-leg $t$-$J$ ladders with $n$ up
to 6 support the notion that the $t$-$J$ model does not phase-separate in
the physically relevant $J/t$ region \cite{RWS99}. This conclusion disagrees 
with
Green's function Monte Carlo calculations of Hellberg and Manousakis
\cite{HM97} but is
in agreement with other Green's function Monte Carlo calculations with
stochastic reconfiguration by
Calandra and Sorella \cite{CS00}.  These latter authors also find evidence for
long-range superconducting order for dopings $x>0.1$ rather than stripes.
Thus, while questions remain regarding the existence of long-range
superconducting order, stripes, and phase separation,
it is clear from a variety of numerical work that the
Hubbard and $t$-$J$ model have low-lying states that
exhibit a number of features which are seen in
the cuprate materials.

So what can we conclude from this about the basic pairing mechanism?
Certainly, in a general way, the fact that these models exhibit
antiferromagnetism, $d_{x^2-y^2}$-pairing correlations and stripes imply
that these phenomenon can arise naturally in systems with strong
short-range Coulomb interactions in which there is a competition between
kinetic and exchange energies: $d_{x^2-y^2}$ pairs as well as domain walls 
form as the
doped holes locally arrange themselves so as to satisfy the competing
requirements of minimizing both their kinetic energy and the disturbance of
the background exchange interactions.  However, beyond this general
picture,
forced by the minimalist nature of these models which after all have only
$U/t$ or $J/t$ along with the band filling $n$ to parameterize the physics,
there remain a remarkably diverse set of views regarding the underlying
pairing mechanism.  

In part, this diversity reflects the fact that for the parameter regime
appropriate to the cuprates ($U/t \gg 1$ or $J/t < 0.5$) these models
describe a strongly-coupled system which is {\it delicately balanced}.  Thus,
changes in the doping, the lattice or the next-near-neighbor hopping can
alter the nature of the groundstate.  While we believe that this delicacy
in fact provides further evidence that these models contain much of
the essential physics, it poses special problems for the theorist.  One
difficulty with a strongly-interacting system is that when a particular
type of mean-field order is assumed, it has a good chance of being
self-consistently ``found''. That is, strongly-interacting systems can
support a self-fulfilling ansatz, keeping the true nature of the groundstate
hidden. One goal of the numerical work is to avoid this. However, it is
easy to slip back into this framework in interpreting the numerical
results. That is, one's description of ``the mechanism'' may depend upon
one's starting ansatz. The delicate balance of the system suggests that one
must simultaneously treat a variety of channels. In addition, one must ask if
the weak-coupling regime of the Hubbard model is continuously connected to
the strong-coupling regime.  That is, what is the nature of the state out
of which the superconducting state is formed as the temperature is lowered
or at zero temperature as the doping is changed.  Depending upon how one
interprets the limited numerical data or what additional interactions or
degrees of freedom one imagines including, different pairing scenarios arise.  

Here, with this in mind, we'll examine some of the numerical results.
Consider the schematic phase diagram for the 2-leg Hubbard ladder shown in
Fig.~1. Here the average site filling $\langle n \rangle$ is plotted along
the horizontal axis and the ratio of the rung hopping $t_\perp$ to the leg 
hopping
$t$ is plotted along the vertical axis. For the non-interacting system, the
antibonding band is lifted above the bonding band for $t_\perp/t>2$. Thus,
at half-filling, $\langle n \rangle =1$, for large values of $t_\perp/t$
the ladder is simply a typical band insulator with an even number of
electrons per unit cell (2 per rung) filling the bonding band.  
In the absence of the interaction $U$, the bonding and antibonding bands of
the half-filled system would overlap and it would become a metal for 
$t_\perp/t<2$. However, with
$U$ present it remains a spin-gapped insulator at half-filling. 
The important point is that for
$\langle n \rangle=1$, the spin-gapped small $t_\perp/t$ insulating state
is {\it adiabatically} connected to the large $t_\perp/t$ Bloch
band-insulating state.  

The behavior of this system under doping is, of
course, quite different depending upon the size of $t_\perp/t$. In the
shaded 2-band region, doped holes enter as quasi-particles with charge $e$
and spin $s=1/2$. Above a value of $t_\perp/t$ which, for the interacting
system depends upon $U$, the
doped holes enter a single band Luttinger liquid and exhibit spin-charge
separation.  
In the shaded region of the phase diagram, the doped holes form
pairs consisting of superpositions of rung and leg singlets with the
$d_{x^2-y^2}$-like phasing shown in Fig.~2. Here, power-law 
pairing
correlations compete with CDW correlations \cite{DRS92,SRZ94,NWS94,DR96}.  
How should these results be
interpreted? One might have thought that
the 2-leg ladder geometry is ideally suited for an
RVB description.  In the limit in which the rung exchange $J_\perp$ is large,
one can imagine that the groundstate of the undoped ladder contains a set
of rung valence-bond singlets with a spin gap $-\frac{3}{4}\, J_\perp$. 
Upon hole doping, 
these latent
rung singlet pairs are free to move and power law pairing correlations
develop.  In the isotropic case in which the leg and rung exchange
couplings are equal, the undoped groundstate still has a robust spin gap
of order $J/2$ and one can again picture that the pairing correlations arise
when the system is hole-doped and the singlet pairs (shown in 
of Fig.~2) are free to move.  Thus, there is a continuous
transition from the spin-gapped Mott insulating phase to the
$d_{x^2-y^2}$-superconducting phase as the chemical potential exceeds a
critical value.  However, contrary to some of the original RVB
tenents \cite{And97}, there are no spinon excitations in the undoped, 2-leg
ladder and the type of spin-charge separation found in the one-leg system
is absent. In fact, for the 2-leg Hubbard ladder the weak and 
strong-coupling regions appear to be adiabatically connected.

Another view argues that the exchange of short-range antiferromagnetic spin
fluctuations mediate the pairing \cite{SLH86,MP94}.  Indeed,
Monte Carlo calculations \cite{DSxx} of the effective pairing
interaction $V(q,\omega_m)$ show that it resembles the spin
susceptibility $\chi(q,\omega_m)$ as suggested by weak-coupling RPA
calculations.  Fig.~3 shows Monte Carlo results for
$V(q,0)$ and $\chi(q,0)$ versus  $q_x$ for
$q_y=\pi$. As the temperature is lowered and the short-range, spin-spin
correlations develop, the pairing interaction $V$ shown in Fig.~3(a)
exhibits a $q$
and $T$ dependence similar to $\chi$ shown in Fig.~3(b).  This same
behavior is seen on $8\times 8$, 2D clusters \cite{BSW93}
However, because of the fermion sign 
problem
these Monte Carlo calculations are limited to temperatures greater than of
order $J/3$.  Moreover, the simple RPA framework clearly fails to describe
the spin-gapped nature of the undoped ladder and an approach such as the
RNG-bozonization \cite{Fab93,LBF97} method is needed. 
In this approach,
a weak-coupling renormalization group calculation is used which treats
all channels on an equal footing until the dominant couplings emerge and
can be treated by abelian bosonization.  Lin, Balents, and Fisher 
\cite{LBF97}, and Arrigoni and Hanke \cite{AH99}
 have shown that in weak coupling a generic ladder model, including both
longer-range Coulomb interactions and hoppings \cite{AH99}, flows
to a manifold with $SO(5)$ symmetry.  This is the symmetry group originally
proposed by Zhang \cite{Zha97} in which the antiferromagnetic and
superconducting order parameters are combined into a five-dimensional
superspin. For an $SO(5)$ symmetric ladder, the magnon
dispersion about $(q_x=\pi,\ q_y=\pi)$ should be identical to the hole pair
dispersion about $(q_x=\pi,\ q_y=\pi)$. This was shown to be the case for
an explicitly constructed $SO(5)$ ladder \cite{SZH98}.  
However, this need not be the
case for the Hubbard and $t$-$J$ models in the physically relevant region
of parameter space. The point is that the RNG results were obtained in the
weak-coupling regime.  Thus, for example, for the Hubbard ladder, the
question is whether there is room for a sufficient renormalization group
flow to approach the region in which $SO(5)$ provides a useful description
of the low energy properties when $U$ is of order the bandwidth. A similar
question, of course, also arises for the $t$-$J$ ladder. One test of
$SO(5)$ is to compare the magnon and pair dispersions which would be
identical for an $SO(5)$ ladder.  The question is whether they are close so
that $SO(5)$ provides a good starting point for describing the collective
modes of the $t$-$J$ ladder, or whether they are significantly different.
With the open boundary conditions on a $2\times L$ ladder used in the DMRG
calculations, the coefficient of the $q^2$ dispersion measured relative to
$(\pi, \pi)$ for the magnon and $(0, 0)$ for the pair, is given by the
slope of the $L^{-2}$-dependence of the energy $E_m = E_0 (S=1)-E_0
(S=0)$ to add a magnon and the energy to add a pair $E_p = E_0
(n_n=2)-E_0(n_n=0)$, respectively.  As shown in Fig.~4, for a 2-leg $t$-$J$
ladder with $J/t=0.5$ these two slopes differ by a factor of 2. Thus, in
the strong-coupling limit the 2-leg Hubbard model, or here the $t$-$J$
model, does not exhibit an exact $SO(5)$ symmetry.  However, recent work on
the Hubbard model and $SO(5)$ symmetry \cite{ZHAHA99} has shown that in
intermediate to strong coupling, the presence of the on-site $U$ requires
one to implement a projected $SO(5)$ symmetry in which the high-energy
double occupancies, responsible for the Mott-Hubbard gap, are
Gutzwiller-projected out. This restores $SO(5)$ symmetry for static
correlation functions but introduces changes in dynamic ones \cite{HAN*},
such as the differences in the magnon and pair dispersions discussed above.

Turning next to the 8-leg, $t$-$J$ ladder results, the DMRG calculations
\cite{WS98a}
for the doped system find evidence for the formation of charged stripes
separating $\pi$-phase shifted antiferromagnetic regions as illustrated in
Fig.~5. Both site-centered ``one-leg'' and bond-centered, ``two-leg''
stripes with a linear charge density $\rho_L=0.5$ have been found in these
calculations, depending upon the length of the open end $8\times L$ system.
While the existence of striped domain walls was found in early Hartree-Fock
calculations \cite{ZGPRS89}, Emery and Kivelson \cite{EK94} have argued that 
stripes 
arise when
phase separation in the $t$-$J$ model is frustrated by long-range Coulomb
forces.  In addition, they have suggested that the holes on the stripes
exhibit charge-spin separation, at least over some suitable length scale,
so that spinons can fluctuate into the ``insulating spin-gapped'' regions
between the stripes.  In this spin-gap proximity effect scenario, these spinons then transfer the 
spin-gap from the
``insulating spin-gapped'' region back onto the charged stripe
leading to enhanced pairing correlations \cite{EKZ97}.
Ultimately the pairing correlations on different stripes couple through a
Josephson coupling to form the superconducting state.

The DMRG calculations
certainly do find striped domain walls in a variety of $n$-leg
ladder calculations.  Furthermore, there is a significant spin gap of order
$J/2$ on an undoped 2-leg ladder which could make such regions attractive
candidates for inducing a ``spin-gap proximity effect''.  Nevertheless,
our numerical calculations on the $t$-$J$ model imply that it does not
phase-separate in the physically relevant region of parameter space
\cite{RWS99}.
Rather, we find that stripe formation in the $t$-$J$ model is driven
by the competition between the kinetic and exchange energies.  Furthermore,
when we turn on a next-near-neighbor hopping $t^\prime$, the stripes and
stripe-stripe correlations decrease as the pairing correlations increase
in strength. The pairing correlations are present when the static stripes
have disappeared as well as any evidence for the fluctuation stripe
correlations in the density-density correlation function.  It appears from
these calculations that
in this region the stripes have {\it evaporated} \cite{WS99}. 

Still other theories propose additional order parameters such as the
$d$-wave density state which has antiferromagnetic orbital currents.  
Monte Carlo calculations on a Gutzwiller
projected BCS $d$-wave variational state are reported to have
a power law decay of the
orbital currents 
 \cite{ILW00}. It has also been suggested that the pseudogap 
phase of the
cuprates is associated with a $d$-wave density phase which is prevented
from fully developing by disorder \cite{CLMN00}.  However, we find that the
antiferromagnetic orbital currents decay exponentially on the 2-leg $t$-$J$
ladder \cite{ASWup} and also decay rapidly on $6$-leg $t$-$J$ ladders.
In addition, were a phase transition to a $d$-wave density
phase to occur for the lightly-doped 2D Hubbard model at a temperature
$T^*$ well above the superconducting transition temperature, one would have
expected to see some indication of this in the finite temperature Monte Carlo
results.

There has also been a recently proposed $Z_2$ gauge theory in which 
electrons can fractionalize into separate charge and spin degrees of
freedom in dimensions greater than one \cite{SF99}. 
Here the spin of the electron is
carried by a neutral fermionic excitation, the spinon, and the charge is
carried by a bosonic excitation, the chargon.  Central to this
fractionalization is an underlying topological order. 
If this $Z_2$
topological order is present, the groundstate of such a system on a 
2D periodic lattice
should exhibit a 4-fold degenerate groundstate. Here, we comment on the
relationship of such a theory to what is known numerically about the 2D
Hubbard and $t$-$J$ models. For the 2D Heisenberg model with periodic
boundary conditions, exact diagonalization calculations on finite sized
clusters find a ``Neel tower of states'' with the groundstate energy $E_0$
varying with the total spin as $S(S+1)$. This implies that the groundstate
of the half-filled $t$-$J$ model is not a ``topological antiferromagnetic''
state which would have had an $S=0$, 4-fold degenerate groundstate in the 
limit of
a large 2D lattice with periodic boundary conditions. Similarly, in the 
weak-coupling limit of the half-filled 2D Hubbard
model, renormalization group calculations imply that the system has a Neel
antiferromagnetic groundstate and electron-like quasi-particles above a
gap.  Since, in strong coupling, the Hubbard
model goes to the $t$-$J$ model,  which at half-filing is Heisenberg-like,
it seems likely that the groundstate of the half-filled Hubbard model is
the traditional antiferromagnetic state for all $U/t$. It may be that
additional 4-site ring exchange interactions, or other interactions laying
outside the simple near-neighbor exchange-coupled Heisenberg model, 
can lead to such a topological
state. This remains to be studied. 

For the doped 2D Hubbard and $t$-$J$ models, it is possible that a
topological phase exists.  This would, of course, be a key finding and
would lay the groundwork for a spin-charge fractionalization mechanism in
which the electron separates into a spinon which carries the electron's
spin and fermi statistics and a chargon which can bose condense giving
superconductivity \cite{SF99}. While it seems unlikely that this can happen at
infinitesimal doping, it could be that it occurs at some finite doping.  At
present, our numerical work on the Hubbard and $t$-$J$ models has failed to
find evidence of such a topological state.  However, these studies are
limited and there are of course additional interactions present.  For example, 
we know that the cuprates are actually charge transfer insulators rather than
Mott-Hubbard insulators.  While we believe that this enhances the pairing
tendencies that one sees in the Hubbard and $t$-$J$ models due to the extra
exchange paths and increased pair mobility \cite{DSW00}, it is possible that 
these
additional interactions produce more profound effects.

To summarize, the Hubbard model and its strong-coupling limit, the $t$-$J$
model, have certainly proven ``instructive'' as orginally suggested by
Gutzwiller.  The numerical results exhibit a
variety of antiferromagnetic, $d_{x^2-y^2}$-pairing and stripe correlations
reminding us of the phenomena seen in the cuprates and providing tests for
various pairing scenarios.  The numerical results suggest that (1) the
competing requirements of minimizing the kinetic energy and the disturbance
of the exchange energy background underlie {\it both} pairing and stripe domain
wall formation, (2) the momentum and temperature dependence of the
effective pairing is {\it similar} to the spin susceptibility, and (3)
stripes {\it compete} with superconductivity.

\acknowledgments

We are grateful to Leon Balents and Matthew Fisher for useful discussions
and insights. D.J.~Scalapino acknowledges support from the NSF 
under grant \# DMR98-17242.  S.R.~White acknowledges support from the NSF
under grant \# DMR98-70930.

\begin{figure}
\centerline{\epsfysize=2in \epsfbox{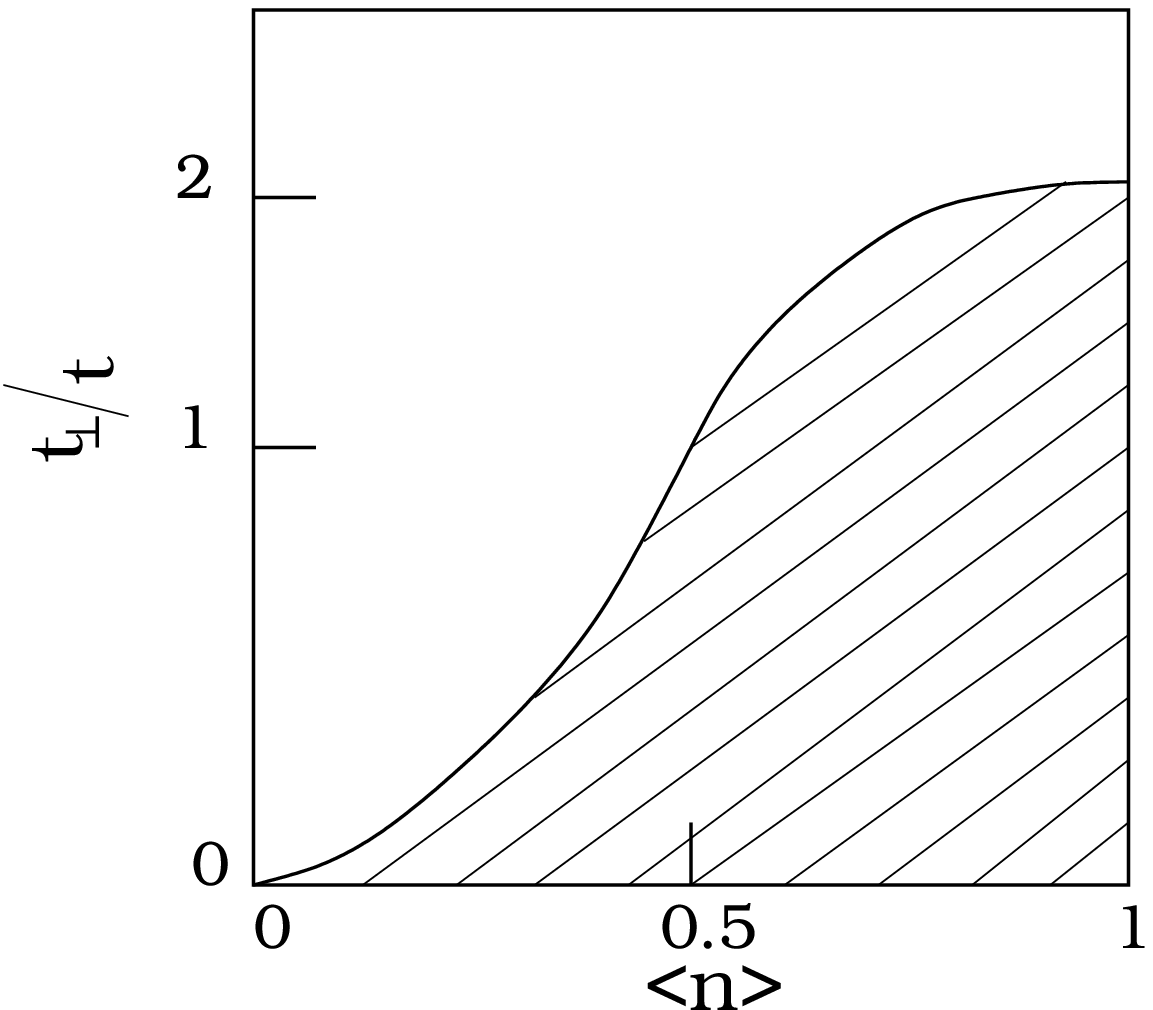}}
\vskip .13 in
\caption{Schematic groundstate phase diagram of the 2-leg Hubbard model
showing $t_\perp/t$ along the vertical axis and the average site filling
along the horizontal axis.  For $U=0$, the solid curve separates the region 
in which
the bonding and antibonding bands both have electrons (shaded)
from the region in which the electrons only occupy states in the bonding
band.}
\end{figure}

\begin{figure}
\centerline{\epsfysize=1.5in \epsfbox{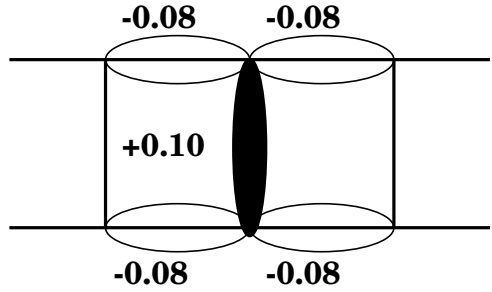}}
\vskip .13in
\caption{Schematic drawing of the pair-wave function showing the values of
the off-diagonal matrix element $\langle N-2 | (c_{i\uparrow}
c_{j\downarrow} - c_{i\downarrow} c_{j\uparrow}) | N \rangle$ for removing
a singlet pair between near-neighbor sites.}
\end{figure}

\newpage
\begin{figure}
\centerline{\epsfxsize=5.5in \epsfbox{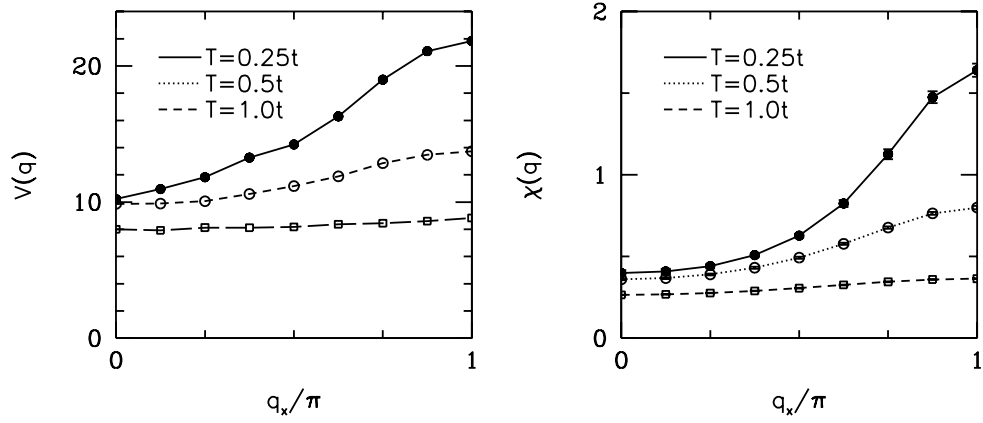}}
\vskip .13in
\caption{(a) Momentum dependence of the effective interaction $V(q)$ at
various temperatures for
a 2-leg Hubbard ladder with $U=4t$, $\langle n \rangle=0.875$ and 
$t_\perp=1.5t$. Here, $V(q)$ is
measured in units of $t$, $q_y=\pi$ and $V(q)$ is plotted as a function of
$q_x$. (b) Momentum dependence of the magnetic susceptibility $\chi(q)$ for
the same parameters.}

\end{figure}

\begin{figure}
\centerline{\epsfxsize=3.0in \epsfbox{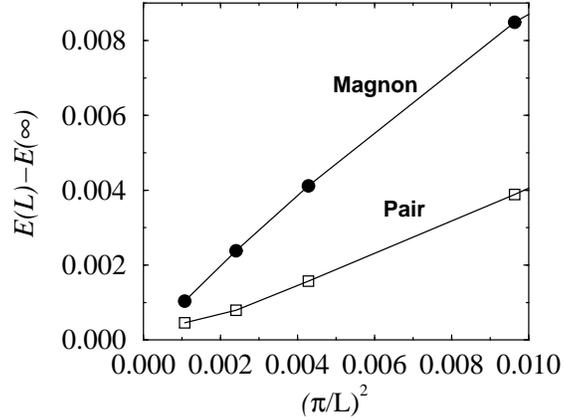}}
\vskip .13in
\caption{Magnon and hole-pair energies versus $(\pi/L)^2$ for a $2\times
L$, $t$-$J$ ladder with $J/t=0.5$. Exact $SO(5)$ symmetry would imply that
the slopes of the two curves would be the same.}
\end{figure}

\newpage
\begin{figure}
\centerline{\epsfxsize=4.0in \epsfbox{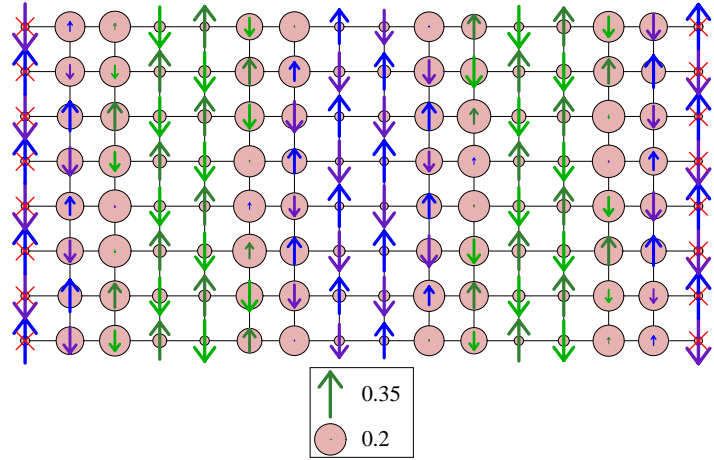}}
\vskip .13in
\caption{Hole density and spin moments for a $16\times 8$, $t$-$J$ lattice
with $J/t=0.35$ and a hole doping $x=0.125$. The diameter of the circles is
proportional to the hole density $1-\langle n_i\rangle$ on the $i^{th}$
site and the length of the arrow is proportional to $\langle S^z_i\rangle$
according to the scales shown.  The lattice has periodic boundary
conditions in the $y$-direction and open boundary conditions in the
$x$-direction with a staggered magnetic field of strength $h=0.1t$
applied at the open ends.}
\end{figure}

\end{document}